\documentclass[fleqn,usenatbib]{mnras}

\usepackage{newtxtext,newtxmath}

\usepackage[T1]{fontenc}
\usepackage{ae,aecompl}
\usepackage{multicol}

\usepackage{graphicx}	
\usepackage{amsmath}	
\usepackage{epsfig}
\usepackage{epstopdf}
\usepackage{float}
\usepackage{caption}
\usepackage{subcaption}

\def\be{\begin{equation}}
\def\ee{\end{equation}}


\title[ DM Halos in Voids]{ Voids and Halos in Voids statistics as a probe of the Expansion History of the Universe}

\author[Laya Parkavousi et al.]{
Laya Parkavousi, Hamed Kameli, Shant Baghram\thanks{baghram@sharif.edu}
\\
Department of Physics, Sharif University of
Technology, P.~O.~Box 11155-9161, Tehran, Iran\\\
}

\date{Accepted XXX. Received YYY; in original form ZZZ}

\pubyear{2023}

\begin{document}
\label{firstpage}
\pagerange{\pageref{firstpage}--\pageref{lastpage}}
\maketitle

\begin{abstract}
Structures in the Universe are arranged into the cosmic web. Distributions, statistics, and evolutions of the structures can be used as probes for cosmological models. We investigate the number density of voids and dark matter halos-in-voids in the Excursion Set Theory (EST). We study the Markov and non-Markov frameworks of EST in both spherical and ellipsoidal collapse models. Afterward, we compare the number density of voids and halos-in-voids in the standard $\Lambda$CDM and the reconstructed model. The reconstructed model is a model-independent reconstruction based on background observations. This work explores the effects of the collapse model barrier in the different EST frameworks on the statistics of voids and the statistics of halos-in-voids.
Finally, we find the hint that cosmological models can be distinguished by the number density of halos-in-voids in the $1.0-2.5$ redshift range. The maximum difference is observed in $z\sim1.9$.
\end{abstract}


\section{Introduction}
\label{Sec:1}
The cosmological structures are distributed in the cosmic-web \citep{Bond:1995yt}. These structures are categorized as dark matter halos, filaments, sheets, and voids. The distribution and the statistics of each element are used to constrain the cosmological models \citep{Percival:2009xn, Alam:2016hwk, Camacho:2018mel}. The calculation of the number density of dark matter halos originated from the inspiring idea of \cite{Press:1973iz}. It has led to the approaches such as excursion set theory (EST) \citep{Bond:1990iw,  Zentner:2006vw, Nikakhtar:2016bju} and peak theory \citep{Bardeen:1985tr}. These approaches relate the abundance of dark matter halos in late time to almost linear and Gaussian initial conditions \citep{Cooray:2002dia}. The same ideas are used to develop the methods to predict the number density of filaments \citep{Fard:2018dwx} and voids \citep{Sheth:2003py, Paranjape:2011bz, Jennings:2013nsa}.\\
Over the last three decades, many developments have been made in the excursion set theory to make it a more realistic semi-analytical model using ellipsoidal collapse  \citep{Sheth:1999mn, Sheth:2001dp}. Furthermore, a sophisticated modification is considered \citep{Maggiore:2009rv, Maggiore:2009rw, Paranjape:2012ks,  Musso:2012qk, Musso:2013pha ,Nikakhtar:2018qqg, Musso:2019zmr, Baghram:2019jlu} to predict the distribution \cite{Tinker:2008ff, Courtin:2010gx}, clustering \citep{Sheth:1999mn}, and merger history \citep{Lacey:1993iv} of cosmic structures, even in alternative models \citep{Lam:2012fa}.\\
A step further would be to study the statistics of dark matter halos in different environments. A more complicated and ultimate test of cosmological models is performed by these probes. This work proposes that dark matter halos (corresponding to galaxy distributions) in voids are a promising tool for testing cosmological models viability  \citep{Tavasoli:2012ai}. Several studies have investigated the distribution, statistics, and characteristics of dark matter halos and galaxies, especially in the underdense region \citep{Voivodic:2020fxt, Tavasoli:2021reo}. According to our proposal, deviations from the standard distribution of haloes/galaxies may have a cosmological origin.\\
We should note that relating the distribution of DM halos and cosmic voids extracted from semi-analytical models such as EST to N-body simulations and observations is a challenging and complicated task, as explored in recent studies for standard model \cite{Furlanetto:2005cc, Shandarin:2005ea, Wojtak:2016brz}, dark energies \cite{Biswas:2010ey, Pisani:2015jha} and even in modified gravitates \cite{Clampitt:2012ub, Voivodic:2016kog, Perico:2019obq}. The main issue with using EST-like models as cosmological probes are connecting analytical models to simulations and observations, which, in some studies, this issue has been addressed \cite{Zhang:2013lda}. The determination of the environment in simulations and observations is also a major obstacle for this study \cite{Sutter:2013pna, Nadathur2015}. In particular, finding cosmic voids is a challenging task \cite{Sutter:2012wh, Sutter:2014kda}. Therefore, this work should be considered as a toy model and a primary suggestion for using these ideas as cosmological probes. Undoubtedly, more robust modeling, observation, and study of the interconnection between semi-analytical models and simulation/observations are needed.\\
There has been a recent challenge to the standard model caused by two separate measurements of $H_0$ from cosmic microwave background (CMB) radiation \citep{Riess:2019cxk} and nearby supernovae \citep{Riess:2019cxk}. Additionally, it provides a new arena for testing models beyond the standard $\Lambda$CDM to reconcile the tension, ranging from early universe models \citep{Poulin:2018cxd} to the late time solutions \citep{Bull:2015stt, Khosravi:2017hfi, DiValentino:2019ffd, DiValentino:2021izs}. A thorough review of the new proposed theoretical models is presented in \cite{DiValentino:2021izs}.\\
The large-scale structure (LSS) is used to test these cosmological models, which address the Hubble tension and the cause of the accelerated expansion of the Universe \citep{Baghram:2010mc, Klypin:2020tud}. This work, with its specific point of view, continues the study by \cite{Kameli:2020kao}, which investigated the effects of models that designed to reconcile the $H_0$ tension on LSS dark matter halo number density and mass assembly history. We focus on the number density of voids and the halo-in-voids as a complementary study and a cosmological probe. 
Specifically, we propose a new probe to distinguish between a phenomenological model with a reconstructed Hubble parameter and a standard $\Lambda$CDM. {{Meantime, we try to address the long-standing challenge of void phenomena in this context \citep{Peebles:2001nv}}}.\\
The paper is structured as follows: In Sec. \ref{Sec:2},  we review the expansion history and its relation with structure formation both in the linear and non-linear regimes. In Sec. \ref{Sec:3}, we review the theoretical background for studying the number density of voids and dark matter halo-in-voids in the context of EST. In Sec. \ref{Sec:4}, we present our results, and in Sec. \ref{Sec:5}, we discuss our conclusions and future directions.\\
\section{ Expansion history and non-linear structure formation}
\label{Sec:2}
\begin{figure} 
	\includegraphics[width=8.5cm]{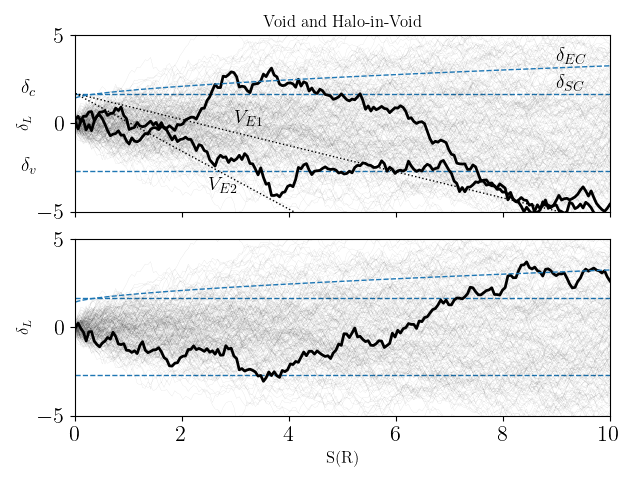}
	\caption{A pedagogical plot of the Excursion set theory in the 2D plane of linear density versus variance. This plot shows the idea behind counting the number density of voids (first down-crossing) and halo-in-voids (an up-crossing after a down-crossing). The SC and EC barriers are shown by a dashed line pointed by $\delta_{SC}$ and $\delta_{EC}$. Also, two inclined dotted lines show the relation of the linear density to variance each for a specific Eulerian radius of voids $V_{E1}$ and $V_{E2}$.} \label{fig2}
\end{figure} 
The purpose of this section is to illustrate the main idea beyond the linear structure formation and the relation between it and the expansion history. Then we review the EST as a non-linear model in both contexts of the Markov and non-Markov extensions of the EST. Additionally, ellipsoidal collapse (EC) and sphere collapse (SC) models are studied.\\
In linear regime, the equations of continuity, Euler and Poisson give rise to the dynamical differential equation for the growth function $D(z)$, which shows the growth of density contrast $\delta\equiv \rho /\bar{\rho}-1$ normalized to unity in present time, and $\delta(z) = D(z) \delta_0$, where $\delta_0$ represents the current density contrast. Where $\bar{\rho}$ is the background density of matter. The dynamical equation is as follows \cite{Kameli:2020kao}
\be \label{eq:growth}
\frac{d^2\delta}{dz^2} + [\frac{dE(z)/dz}{E(z)}-\frac{1}{1+z}]\frac{d\delta}{dz} - \frac{3}{2}\Omega_m\frac{1+z}{E^2(z)}\delta = 0,
\ee
where $E(z)=H(z)/H_0$ represents the normalized Hubble parameter and $\Omega_m$ is the present time matter density parameter.
The variance of the matter perturbation is related to the linear matter power spectrum at present time $P_L(k,z=0)$ as
\be \label{eq:variance}
S(R)\equiv\sigma^2(R)=\frac{1}{2\pi^2}\int dk k^2P_{\text{L}}(k,z=0)\tilde{W}^2(kR),
\ee
where $\tilde{W}(kR)$ is the Fourier transform of the smoothing window function and the matter power spectrum is
\be
P_{\text{L}}(k,z) = A_l k^{n_s}D^2(z)T^2(k),
\ee
Here we use the \cite{Eisenstein:1997ik} transfer function $T(k)$ and $A_l$ is the late time amplitude of the power spectrum normalized by $\sigma_8$.\\
In equation \ref{eq:growth}, the expansion history of the Universe acts as the friction term that affects the growth of the structures. In the standard model, this term is determined by matter density and the present-time Hubble constant. The alternative models of cosmologies have their own expansion histories, which affect the growth of the structures \citep{Baghram:2010mc}. \\

A Hubble tension of nearly $\sim 4-5\sigma$ has been found by recent observations of the SNe-Ia and Planck collaborations \citep{Riess:2019cxk}. In addition to many theoretical alternative models for relaxing this tension, there is an idea of a model-independent reconstruction of $H(z)$ to show the deviation from the standard model prediction. Our model-independent approach describes how late-time expansion history affects voids and halo-in-voids statistics.\\
This is accomplished by using an alternative model for the Universe's expansion history\cite{Wang:2018fng}.  This model-independent approach is used to construct the Hubble parameter based on background cosmological data.
 \cite{Wang:2018fng} used SNe Ia -Joint Light Analysis (JLA) \cite{Wang:2018fng}, SNe Ia -Nearby local measurements \cite{Riess:2016jrr}, BAO -6dF Galaxy Survey (6dFGS) \cite{beutler33666df}, BAO-SDSS DR7 Main Galaxy Sample (MGS)  \cite{ross2015clustering}, BAO -tomographic BOSS DR12 (TomoBAO) \cite{Wang:2016wjr}, BAO - Lyman-$\alpha$ of BOSS DR11 \cite{Font-Ribera:2013wce},  BAO - Lyman-$\alpha$ of BOSS DR11 \cite{Delubac:2014aqe}.\\
This paper investigates the effect of the modified Hubble parameter and growth function (for more discussion see Fig.(1) and Fig.(3) of \cite{Kameli:2020kao}) on the LSS observables in a non-linear regime such as void distribution and halo-in-voids. We use EST as a framework for this task. One should note that the errors in the reconstructed expansion history are not Gaussian. Although, we use a Gaussian approximation for the errors in reconstructed expansion history, which propagates in growth function and EST functions with Gaussian $\pm1\sigma$ error. \\
In EST, the density contrast in each point of the initial field versus variance executes a random walk for a k-sharp smoothing window function.
The number density of dark matter halos is determined by statistics of the first up-crossing of the Markov random walks from a specific barrier which is related to the collapse model. For the SC, with a constant barrier of $\delta_c$, the first up-crossing $f_{\text{FU}}$ of Markov random walks is as below
\be
f_{\text{FU}}(S,\delta_c)dS=\frac{1}{\sqrt{2\pi}}\frac{\delta_c}{S^{3/2}}\exp[-\frac{\delta_c^2}{2S}]dS.
\ee
The number density of dark matter halos is:
\be
n(M,t)dM=\frac{\bar{\rho}}{M}f_{\text{FU}}(S,\delta_c)|\frac{dS}{dM}|dM.
\ee
Note that the number density of dark matter halos can be calculated at any redshift. In this case, the collapsing barrier is modeled as a redshift-dependent function $\delta_c(t)=\delta_c / D(z)$. It is interesting to note that the redshift dependence of the barrier is controlled by the growth function, which depends on the expansion history of the Universe. In the next section, we will discuss the {\it{two barriers}} problem in EST and the physics of void statistics.\\
\begin{figure*}
	\includegraphics[width=17.5cm]{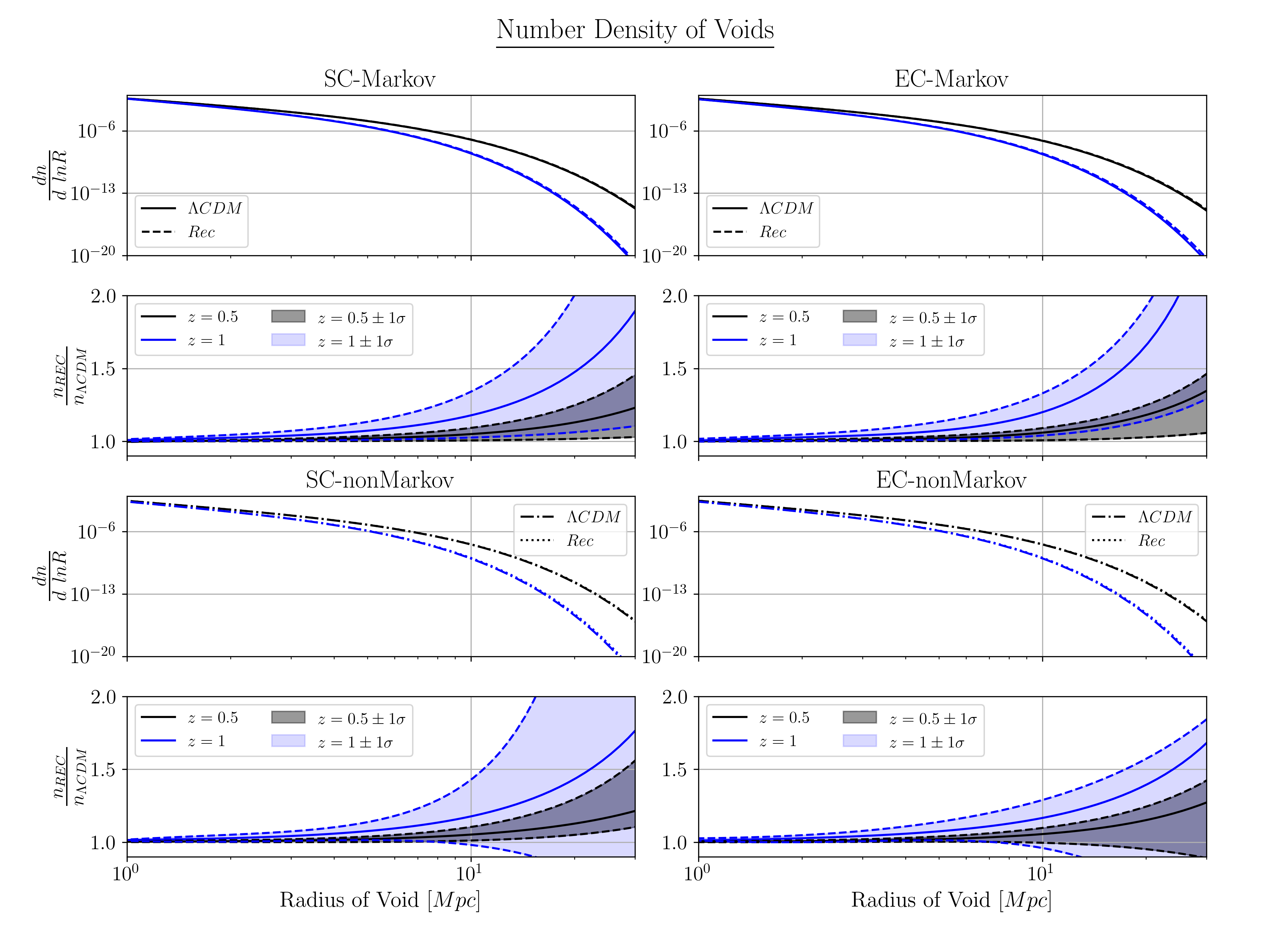}
	\caption{The number density of voids and their ratio in two cosmological models (reconstructed model (Rec) to $\Lambda$CDM) are plotted versus Eulerian radius. In the left column, we show the spherical collapse (SC) model and in the right column, the results are obtained for ellipsoidal collapse (EC). The upper two rows are for the Markovian trajectories and the lower two rows are dedicated to non-Markov trajectories. In all plots, we show the prediction of the $\Lambda$CDM and reconstructed model (Rec) in two redshifts $z=0.5$ (black lines) and $z=1.0$ (blue lines).  The shaded regions in ratio plots show the 1-$\sigma$ confidence level corresponding to the propagation of the error from the ratio of the reconstructed Hubble parameter to $\Lambda$CDM. } \label{fig:nVoid}
\end{figure*}
\begin{figure*}
	\centering
	\begin{minipage}[t]{0.475\linewidth}		
		\centering
		\includegraphics[width=9cm]{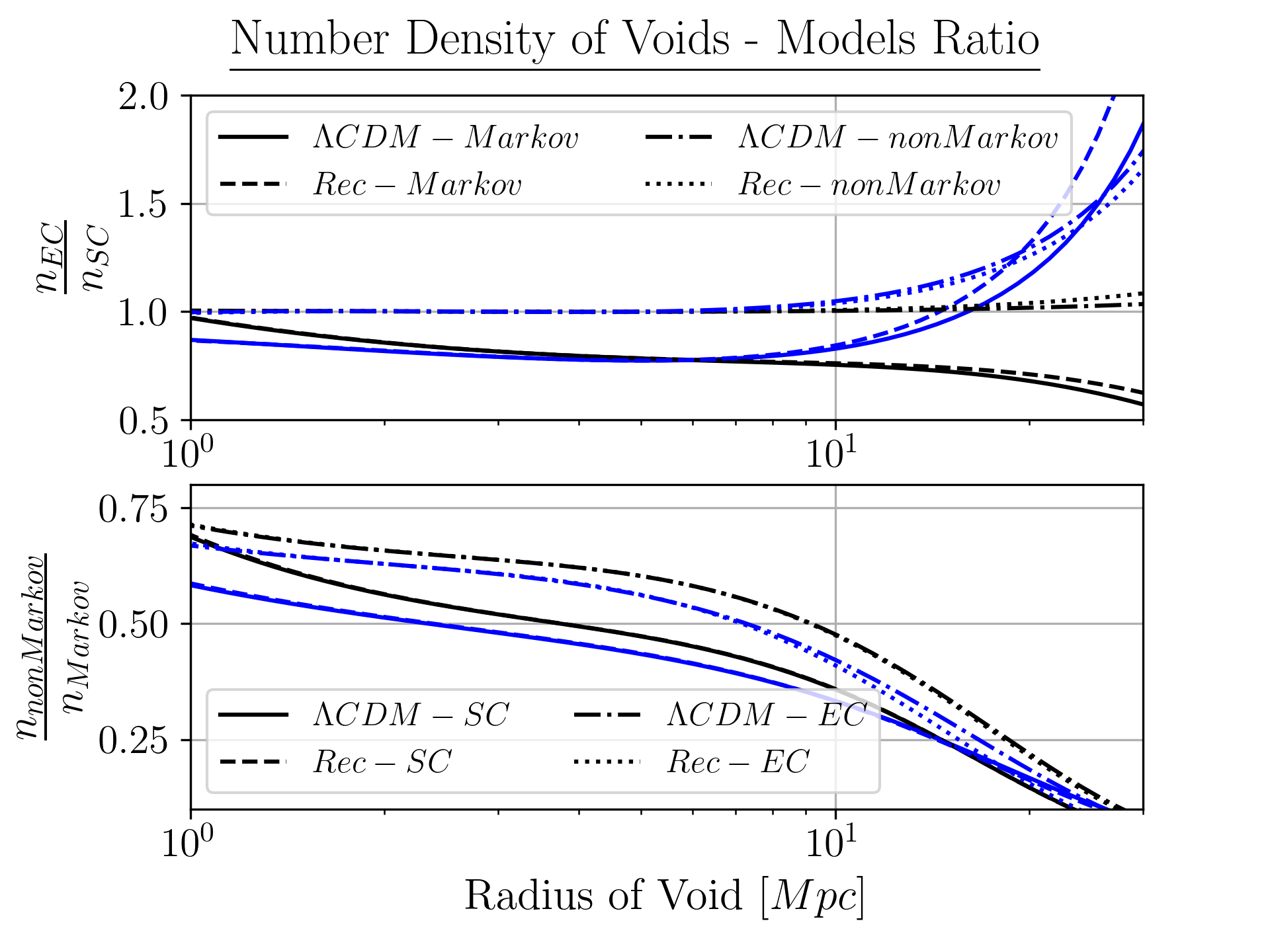}
		\captionof {figure} {Upper panel: The ratio of the number density of voids in EC to SC is plotted versus Eulerian radius. These ratios are shown for different cosmological models ($\Lambda$CDM and Rec) and also Markov and non-Markov trajectories. Bottom panel: The ratio of the number density of voids in non-Markov to Markov trajectories is plotted versus Eulerian radius. These ratios are shown for different cosmological models ($\Lambda$CDM and Rec) and different collapse models (SC, EC). In all plots, the black lines show $z=0.5$ and the blue lines show $z=1.0$.}
		\label{fig:nVoid-ratio}
	\end{minipage}%
\hspace{0.7cm}
	\begin{minipage}[t]{0.475\linewidth}
		\centering
		\includegraphics[width=9cm]{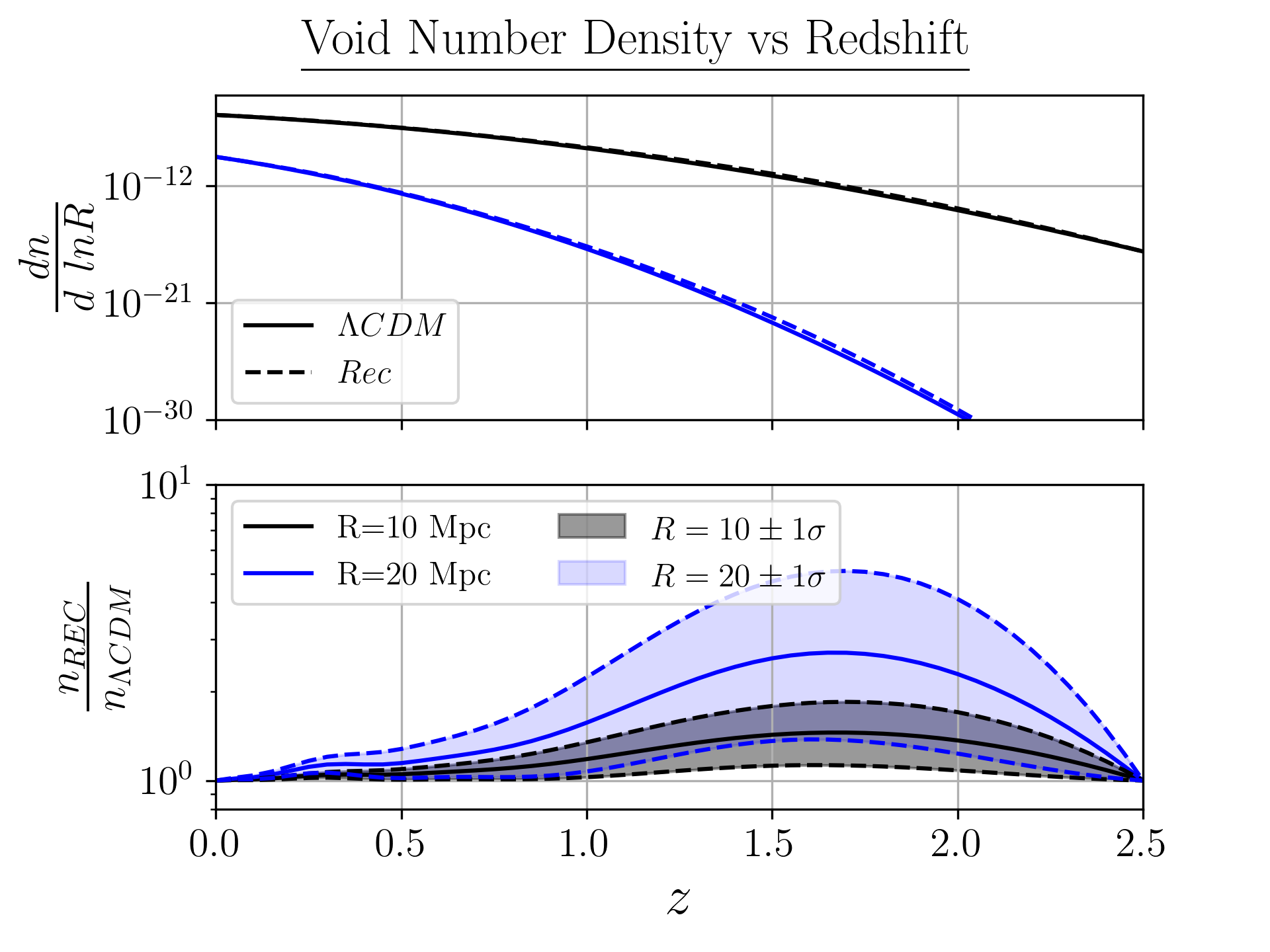}
		\captionof {figure} {The number density of voids for SC-Markovian framework are plotted versus redshift for two specific Eulerian radii $R=10,20$ Mpc with black and blue lines respectively. The upper panel shows the prediction of the two cosmological models ($\Lambda$CDM and Rec) and the bottom panel shows the ratio. The shaded regions show the 1-$\sigma$ confidence level. See App. \ref{app:II} for the other void radii.}
		\label{fig:nVoid-z}
	\end{minipage}
\end{figure*}
\section{ Voids and Halos-in-voids in context of EST}
\label{Sec:3}

According to \cite{Sheth:2003py}   influential work, the abundance of voids is derived from the two barriers EST. The number of voids is related to the number of first down-crossings of a specific barrier $\delta_v$, associated with the shell-crossing barrier $\delta_v\simeq -2.7$. In contrast, the number density of dark matter halos in voids can be determined from trajectories, which have an up-crossing in larger variance after a void formation in smaller variance. As seen in Fig. (\ref{fig2}), we show the schematic configuration for a trajectory with a first down-crossing and then an up-crossing.
According to \cite{Sheth:2003py}, the fraction of the walks that first cross $\delta_v$ at $S$ without crossing $\delta_c$ is calculated as follows.
\be\label{eq:ffuvoid}
S{\cal{F}}(S,\delta_v,\delta_c) = \sum_{j=1}^{\infty} \frac{j^2\pi^2{\cal{D}}^2}{\delta_v^2/S}\frac{\sin(j\pi{\cal{D}})}{j\pi}\exp\left(-\frac{j^2\pi^2{\cal{D}}^2}{2{\delta_v}^2/S}\right),
\ee
where
\be\label{eq:Dvoid}
{\cal{D}}\equiv \frac{|\delta_v|}{(\delta_c + |\delta_v| )}.
\ee
Now we can relate the number density of voids with mass $m$ to the statistics of first down-crossing as follows:
\be
\frac{m^2n_v(m)}{\bar{\rho}}=S{\cal{F}}(S,\delta_v,\delta_c) \frac{d\ln S}{d\ln m},
\ee
where $\bar{\rho}$ is the background comoving density. The mass conservation indicates that the non-linear overdensity $\delta_{\text{NL}}$ is defined as
\be
1+\delta_{\text{NL}} = \frac{m}{\bar{\rho}V_{\text{E}}} \approx (1-\frac{\delta_\text{L}(t)}{\delta_c})^{-\delta_c}, \label{eq:NL-L}
\ee
where $V_{\text{E}}$ is the Eulerian volume. As shown in Fig.(\ref{fig2}), the two dotted lines indicated by $V_{E1}$ and $V_{E2}$ illustrate the relationship between the fixed Eulerian volume to $\delta_L$ through equation \ref{eq:NL-L}.
There has been a discussion in \cite{Sheth:2003py} about studying halo-in-voids based on the EST theory. A halo in a void is represented by trajectory paths that first up-cross the collapsing barrier after first down-crossing the barrier of the voids. In Fig.(\ref{fig2}), the bottom panel, we show a sample trajectory of halo-in-void. It is important to note that the variance related to down-crossing is smaller (related to the larger volume) than the variance of up-crossing (related to the smaller volume/mass) of the halo. In the EST plane, we plot the density contrast versus variance, which is related to the Lagrangian (initial) radius $R_L$. Considering mass conservation, equation \ref{eq:NL-L}, and the non-linear void density contrast $\delta_{{\text{NL,v}}}= -0.8$, will result in $R_E/R_L \simeq 1.7$, where $R_E$ is Eulerian radius. The radius of voids will be Eulerian in the following section. \\
Conditional crossing statistics can be used to determine the density of halo-in-voids. In the case of Markov with spherical collapse of halos \cite{Gunn1972}, this can be obtained \citep{Mo2010}
\begin{eqnarray} \label{eq:ffuc}
f_{\text{FU}}(S_h(M),\delta_c(z)|S_v(R),\delta_v(z))=\frac{1}{\sqrt{2\pi}}.\frac{\delta_c(z) - \delta_v (z)}{(S_h(M)-S_v(R))^{3/2}} \\ \nonumber \times\exp\left({-\frac{(\delta_c(z) - \delta_v(z))^2 }{2(S_h(M)-S_v(R))}}\right),
\end{eqnarray}
Where $S_h(M)$ and $S_v(R)$ are halo and void variances, respectively. The spherical collapse barrier and the void formation barrier are represented by $\delta_c(z)$ and $\delta_v(z)$, respectively.
The number density of voids and halo-in-voids in the ellipsoidal collapse barrier is also calculated, which is a more realistic model for collapse due to N-body simulations for smaller mass DM halos \cite{Sheth:1999mn, Sheth:2001dp, Robertson:2008jr}.
The ratio of critical density in ellipsoidal $\delta_\text{ec}$ to spherical collapse $\delta_\text{sc}$ due to \cite{Sheth:2001dp} model is as below:
\begin{equation} \label{eq:ellip-barrier}
\frac{\delta_\text{ec}}{\delta_\text{sc}} = \sqrt{\bar{a}} [1+\beta (\bar{a}\nu)^{-\alpha}],
\end{equation}
where $\bar{a}\approx 0.7$, $\alpha \approx 0.615$ and $\beta \approx 0.485$.
It is important to note that an analytical approach to the conditional probability function will not be applicable to the ellipsoidal case.
Therefore, we have to use numerical methods to count the crossing statistics.
The numerical method involves generating trajectories and counting up-crossings and down-crossing statistics numerically. For this task, we generate $10^8$ trajectories for both the $\Lambda$CDM and the reconstructed models.\\
For the non-Markov case of trajectories which was raised due to a more realistic smoothing function in the EST framework, we use the methods developed in \cite{Nikakhtar:2018qqg, Kameli:2019bki}. Compared to Markov plots, non-Markov trajectories produce smoother trajectories since the height of the smoothed density field extrapolated to the present time is correlated to the density contrast in all the previous variance steps. A more comprehensive discussion can be found at \citep{Nikakhtar:2018qqg, Baghram:2019jlu, Kameli:2019bki}.\\
In the non-Markov case, for finding the number density of voids we produce $10^8$ trajectories which are memory dependent. Then we count the first down-crossings from $\delta_v$ by excluding the trajectories representing the void in void and void in halos \cite{Sheth:2003py}. For halos-in-voids, the process is much more complicated. First, we should choose the trajectories that touch the $\delta_v$ in a specific voids radius (host void) and then find the statistics of the first up-crossing of the remained trajectories, which show the mass of the embedded halo in the host void.
Crossing statistics obtained by numerical methods are noisy. For this purpose, we used a curve-fit smoothing algorithm to smooth the noisy numbers density of voids and halo-in-voids. By comparison, the curve-fit error is less than $\sim2\%$, whereas the reconstruction method introduces larger error bars.  
For the first time, we propose non-Markov solutions for voids statistics for both spherical and ellipsoidal collapses in section \ref{Sec:4}. Furthermore, we investigate the number density of dark matter halo-in-voids in SC and EC.\\

\section{Results}
\label{Sec:4}

{{
We present our findings in this section. To begin with, we present the number density of voids in Markov and non-Markov cases, and in two different collapse scenarios for $\Lambda$CDM and the reconstructed model. \\
In Fig. (\ref{fig:nVoid}), the number density of voids is shown for both cosmological models, Markov and non-Markov trajectories, SC/EC collapse model at two redshifts $z=0.5$ and $z=1.0$, confirming the hierarchical void formation \citep{Sheth:2003py}.  For all figures, black lines indicate results for $z=0.5$ and blue lines indicate results for $z=1.0$. \\
The number density of voids in EST is calculated by counting the first down-crossing of $\delta_v$ after excluding two types of trajectory. The two are a) void-in-void and b) void-in-halo. In order to exclude void-in-halo trajectories, we need a collapse model for DM halos. In this study, we examine both SC and EC models. An ellipsoidal collapse will result in a variance-dependent critical density.\\
In all models (collapse and cosmological) the number density of voids is less in higher redshifts due to hierarchical structure formation, but the difference between the $\Lambda$CDM and reconstructed models is more significant in $z=1.0$ in comparison to $z=0.5$. This can be explained by the larger difference in the Hubble parameter of the reconstructed model at $z=1.0$.\\
In addition, Fig.(\ref{fig:nVoid}) shows that for void radii $R>10$Mpc the ratio can be exceeded from a couple of percent to a significant value. To give a more reliable conclusion, we calculate the propagation of the error in the number density of voids with a 1$\sigma$ confidence level. The error corresponds to the Hubble parameter reconstruction. Therefore, we show that the two models are indistinguishable by 1$\sigma$ confidence level. To distinguish between different cosmologies with the number density of voids, we need more precise observations to reconstruct the Hubble parameter. The error propagation from the reconstructed Hubble parameter to the number density of voids is monotonically increasing. For more discussion on error propagation see App \ref{app:I}. 
It appears that the number density of voids in two cosmological models follows the same pattern regardless of (Markov, non-Markov) trajectory, and collapse scenario choice. \\
Fig. \ref{fig:nVoid-ratio} shows the ratio of void number density for different collapse models (SC and EC) in the upper panel, and EST models (Markov and non-Markov) in the bottom panel for both cosmological models. Due to smoother non-Markov trajectories in EST, the number density of voids in the non-Markov case is lower than in the Markov one. We plot the number density of voids versus redshift in two cosmological models to determine the appropriate redshift to distinguish the models.\\
We should emphasize that the primary purpose of this work is to compare the two cosmological models and their effects on voids and halos-in-voids statistics. Accordingly, the focus is not on the difference between Markovian and non-Markovian trajectories and the resulting voids statistics. The Figure (\ref{fig:nVoid}) and bottom panel of Figure (\ref{fig:nVoid-ratio}) show the significant impact of non-Markovian trajectories on voids number density statistics. It means the result must be interpreted carefully based on the calculation approach. In the Markovian case, there is a growing bias for void size in the reconstructed model versus the standard model. However, introducing a non-Markovian framework wipes out this bias, with the errors making this trend not statistically relevant. However, we indicated that the ratio of the statistics in the two cosmological models remained the same. \\
\begin{figure*}
	\includegraphics[width=17.5cm]{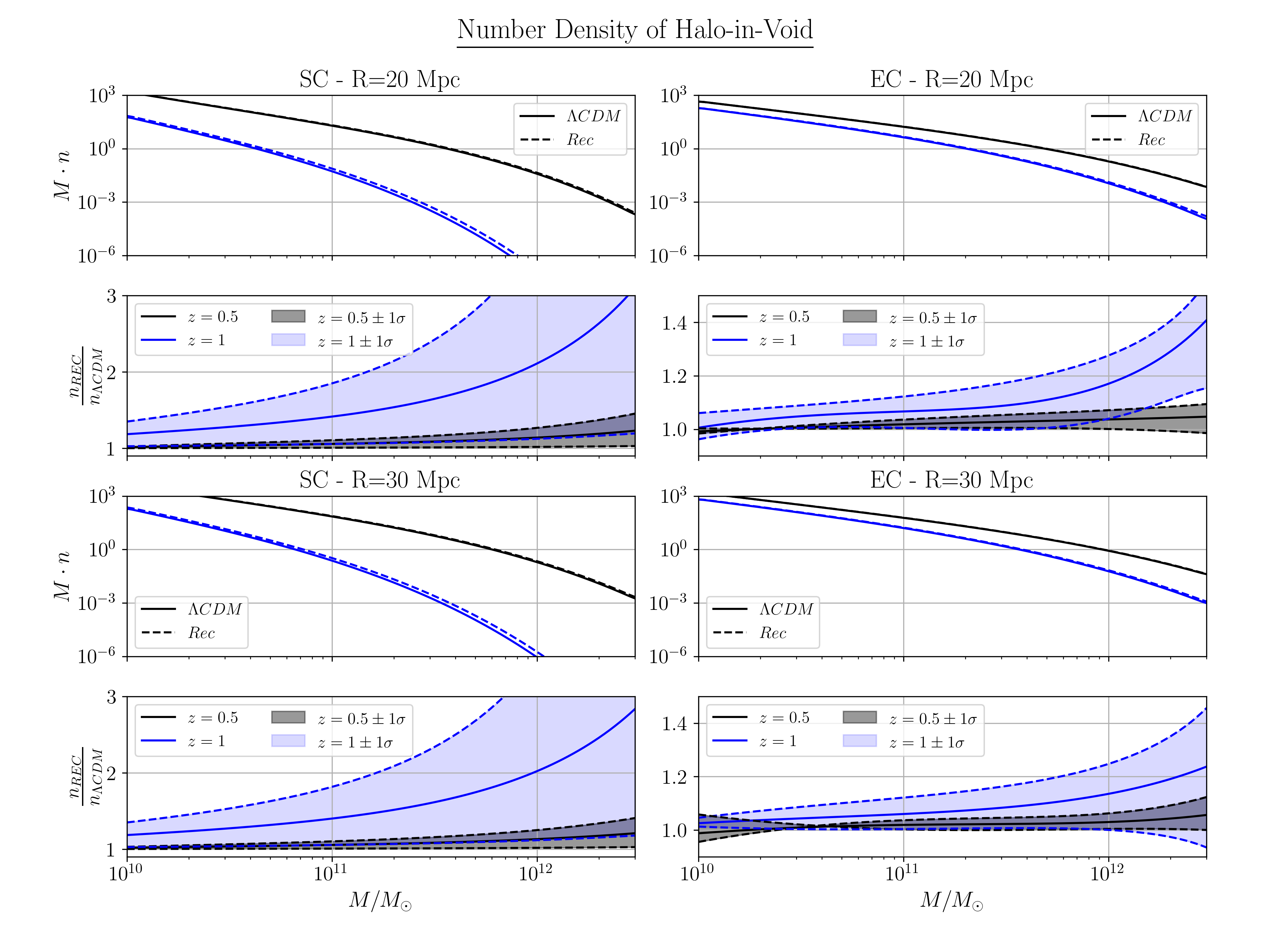}
	\caption{The number density of halos-in-voids and their ratio in two cosmological models (reconstructed model (Rec) to $\Lambda$CDM) are plotted versus DM halo mass. In the left column, we show the spherical collapse (SC) model and in the right column, the results are obtained for ellipsoidal collapse (EC). The upper two rows are for halos embedded in voids with $R=20$Mpc, and the lower two rows are dedicated to the halos in voids with $R=30$Mpc. In all plots, we show the prediction of the $\Lambda$CDM and reconstructed model (Rec) in two redshifts $z=0.5$ (black lines) and $z=1.0$ (blue lines).  The shaded regions in ratio plots show the 1-$\sigma$ confidence level corresponding to the propagation of the error from the ratio of the reconstructed Hubble parameter to $\Lambda$CDM. All the plots are for Markov trajectories.} \label{fig:haloVoid}
\end{figure*}
\begin{figure*}
	\centering
	\begin{minipage}[t]{0.475\linewidth}		
		\centering
		\includegraphics[width=9cm]{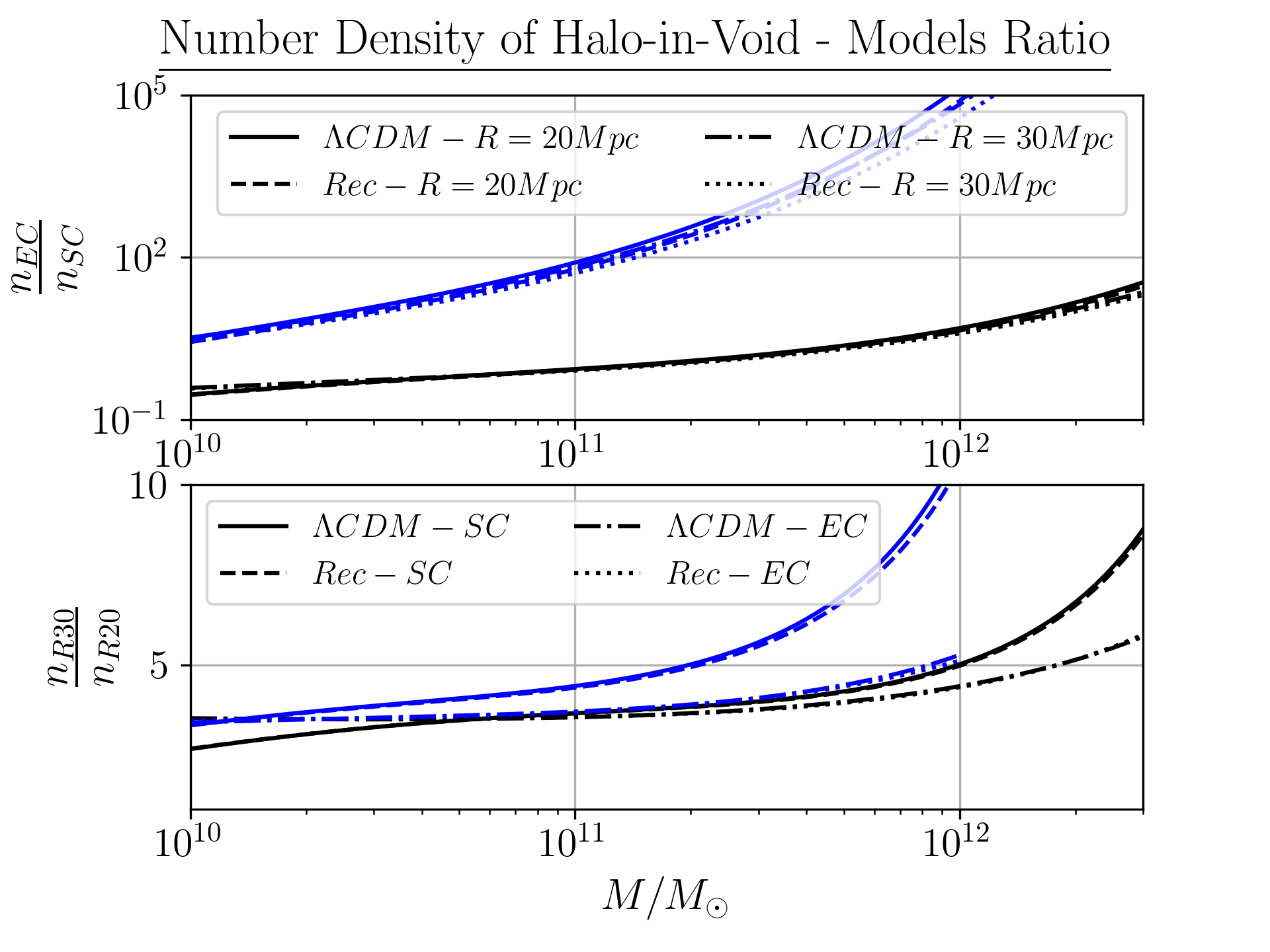}
		\captionof {figure} {Upper panel: The ratio of the number density of halo-in-voids in EC to SC is plotted versus DM halo mass. These ratios are shown for different cosmological models ($\Lambda$CDM and Rec) and also for different host void radii $R=20,30$ Mpc. Bottom panel: The ratio of the number density of halo-in-voids in host voids of radii $R=30$Mpc to $R=20$Mpc is plotted versus DM halo mass. These ratios are shown for different cosmological models ($\Lambda$CDM and Rec) and different collapse models (SC, EC). In all plots, the black lines show $z=0.5$ and the blue lines show $z=1.0$.}
		\label{fig6}
	\end{minipage}%
	\hspace{0.7cm}
	\begin{minipage}[t]{0.475\linewidth}
		\centering
		\includegraphics[width=9cm]{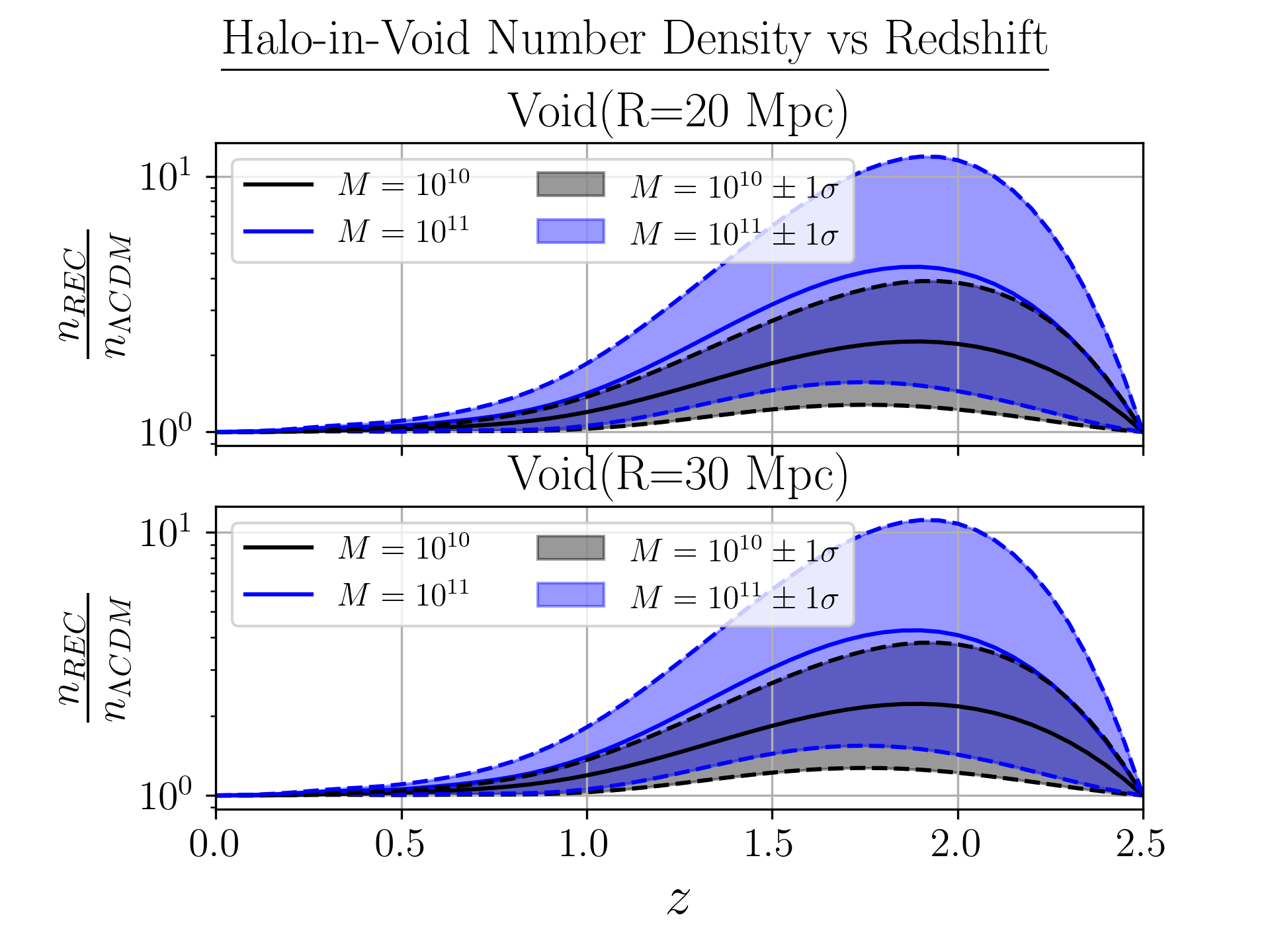}
		\captionof {figure} {The number density of halo-in-voids in SC-Markovian framework are plotted versus redshift for two specific DM halo mass $M=10^{10}, 10^{11} M_{\odot}$  with black and blue lines respectively. The prediction of the two cosmological models ($\Lambda$CDM and Reconstructed plotted for host void $R=20$Mpc radius in the upper panel and $R=30$Mpc in the bottom panel. The shaded regions show the 1-$\sigma$ confidence level. See App. \ref{app:II} for the other halo masses.}
		\label{fig7}
	\end{minipage}
\end{figure*}
In  Fig. \ref{fig:nVoid-z} upper panel, we plot the number density of voids versus redshift for voids with radii $R=10$Mpc and $R=20$Mpc. In the bottom panel, we plot the ratio of the number density of voids versus redshift for different cosmological models and specified radii.  $z\sim 1.7$   is the redshift at which the greatest difference between the two cosmological models can be observed. The number density ratio between two cosmological models versus redshift is correlated with the deviation of the Hubble parameter from the $\Lambda$CDM (see Fig. 3 of \cite{Kameli:2020kao}). A larger difference for $R=20$Mpc confirms the results in the Fig. (\ref{fig:nVoid}).\\
As a proof of concept, we use the SC-Markov model as the only existing analytical solution to find the variation of the void number density with redshift. We do not assert that SC is the optimal model, and we should interpret this result carefully. The non-Markov case and EC model (in which the calculations are done by counting the up-crossings and down-crossings of trajectories) can change the number density of voids. Non-Markovian/Markovian number density is not flat, as shown in Fig. (\ref{fig:nVoid-ratio}). So, it may impact the redshift at which differences are maximal. However, we showed that the main contribution to the void number density ratio comes from the cosmological models, and it is not significantly affected by the collapse model or the Markov or non-Markov schemes. \\
The main contribution of this study is the use of voids and halo-in-voids statistics to distinguish cosmological models. Our next step is to investigate the number density of halos-in-voids. We find galaxies in underdense regions, as discussed in the introductions. Accordingly, the number density of halos in large voids is a fair indicator of the mass (luminosity) distribution of galaxies. \\
In Fig. (\ref{fig:haloVoid}), we plot the number density of halos versus mass in two voids with Eulerian radius of $R=20 {\text{Mpc}}$ and $R=30 {\text{Mpc}}$ in redshifts $z=0.5$ and $z=1.0$  with SC and EC collapse scenario, for two cosmological models with $1\sigma$ error-bars.  We used the counting method for finding the number of first crossings. For larger masses in the EC case, the first-crossing statistics decrease dramatically. And the corresponding errors become large. 
{{Accordingly, in Fig. (\ref{fig:haloVoid}) in the mass range of $M>10^{12}M_{\odot}$ for $z=1$ where the statistics are less than the $z=0$,  the effect of the errors are much emphasized. The error propagation from the reconstructed Hubble parameter to the number density of halos-in-voids is monotonically increasing. For more discussion on error propagation, see App \ref{app:I}.}} It is also worthwhile to clarify here that the computation of conditional crossings with non-Markov trajectories is complicated and computationally expensive. The statistics for conditional crossings are less than the trajectory numbers in non-Markov trajectory types. A linear relationship exists between the number of trajectories and the computational algorithm's running time \footnote{The main problem with the computational cost is the exponential behavior of the number density in large masses. For example, referring to Fig. (\ref{fig:nVoid}), the number density of voids in $R\sim10$ Mpc is $\sim 10^{-8}$, and for $R\sim30$ Mpc, the number density decreases exponentially to $\sim10^{-20}$. The number density is the ratio of the number of crossing to all trajectories. When we want to calculate the ratio of two different cosmological models, the problem is more severe. Also, we have the same problem for $z=1$ in comparison with $z=0$, Due to smaller number density in larger redshift. In our computational approach of counting the first crossing of trajectories (in non-Markov and EC models), for more precision and larger mass range, one needs more trajectories which almost increase the computational cost linearly. } .\\

Furthermore, we want to calculate the conditional probability of halos in voids. These conditional statistics become even more negligible in the non-Markov case (in comparison with number density statistics) in the large mass range. The halos-in-voids condition in the non-Markov case eliminates a significant number of trajectories, which do not satisfy the conditional crossing criteria. This issue leads to several orders of magnitude of computational cost.\\
This problem is primarily caused by conditional crossing statistics. The major computational challenge is not in producing trajectory and non-conditional void statistics (discussed in \cite{Nikakhtar:2018qqg}) but in the conditional counting algorithm. First, we need to extract trajectories with the first down-crossing in small variances (equivalent to larger voids). Following that, we need to find the trajectories in the extracted set which have their first up-crossing in smaller variances (smaller halo masses). In Markov EST, trajectories do not have memory. As a result, we can generate trajectories from any specific point in the EST 2D plane (variance, density contrast), unlike non-Markov EST.  For a specific void size, we must exclude most trajectories that do not touch the voids barrier. Thus, the number of trajectories starting from the void is extremely low. We show this process in Fig. (\ref{fig2}), bottom panel.\\
In order to compare our results with observations, we need halos-in-voids statistics for each radius of the void. However, \cite{Punyakoti:2021mn} proposed an integrated statistic of halos-in-voids which includes the number density of halos in voids, filaments, and nodes.
The plot shows that for large halo masses, the two models deviate from unity. The significant deviation at $z=1.0$ suggests that surveys that probe the structure in underdense regions at higher redshifts are prominent in distinguishing the models. It is also worth noting that the halo-in-void statistics are more sensitive to barrier models. This is an important difference from the statistics of the voids. We then calculate the probability of propagating the error in the number density of voids with a 1$\sigma$ confidence level to make a certain conclusion. The error corresponds to the Hubble parameter reconstruction.\\

The upper panel of Fig. (\ref{fig6}) shows the number density of the halo-in-voids ratio of EC to SC for two different cosmological models and two redshifts $z=0.5$ and $z=1.0$ for different host void radii $R=20,30$ Mpc. As shown in the bottom panel, the relationship between the number density of halo-in-voids in host voids for radii of $R=30$Mpc and $R=20$Mpc is plotted against DM halo mass. We have a significant dependence on the barrier model in the halos-in-voids. In different cosmological models, the main difference is not rising from collapse. The ratio of voids number density and halos-in-voids for two cosmological models is almost independent of the collapse model, and each can be used to distinguish them.\\
In Fig. (\ref{fig7}), we plot the ratio of halo-in-voids versus redshift for host voids with radii $R=20$Mpc (upper panel) and $R=30$Mpc (bottom panel) for two cosmological models. There is a strong correlation between the largest difference in cosmological models, observed in $z\sim 1.9$, and the deviation of the reconstructed Hubble parameter from the $\Lambda$CDM (see Fig. (3) of \cite{Kameli:2020kao}).\\
As a conclusion to this section, we would like to highlight our main finding once again. The number density of voids and halo-in-voids, which is a more complicated observational probe, can be used to test cosmological models that address $H_0$ tension.

\section{Conclusion and Future remarks}
\label{Sec:5}
The Universe's large-scale offers a unique opportunity to test cosmological models to understand dark matter and dark energy. The cosmic LSS are distributed in a web-like structure and categorized as halos, filaments, sheets, and voids. \\
In this manuscript, we study the usefulness of voids and halos-in-voids number densities as a cosmological probe. We seek to identify the optimal redshift at which the deviations from $\Lambda$CDM can be sought. We calculate these number densities using EST for both spherical collapse (SC) and ellipsoidal collapse (EC) collapse models and also consider the impact of non-Markov trajectories. The construction of number densities using EC in the EST methodology is a first study, as the exploration of the non-Markovian trajectories on void number densities. 
We find that the number density of voids is lower at higher redshifts. The number density ratio between two cosmological models versus redshift is correlated with the deviation of the Hubble parameter from the $\Lambda$CDM. $z\sim 1.7$ is the redshift at which the most significant difference between the two cosmological models is observed.
Worths to mention that the ratio can exceed a few percent to a statistically meaningful value in Markovian predictions for void radii $R>10$Mpc.
However, in 1$\sigma$ confidence level, the two models are indistinguishable for void statistics. More precise observations are needed to reconstruct the Hubble parameter to distinguish between different cosmologies.\\
We can distinguish two cosmological models by their halos-in-voids ratio in contrast to the void statistics.
To set the stage, we compare the standard $\Lambda$CDM with the reconstructed one. The reconstructed model is based on a model-independent parametrization of the Hubble parameter with the background observation data. \\
The most consequential difference in cosmological models for halos-in-voids statistics observed in $z \sim 1.9$, which correlates strongly to the deviation of the Hubble parameter from the $\Lambda$CDM.\\
The number density of voids in the non-Markov framework of EST is lower than in the Markov framework. The non-Markov case has smooth trajectories that are memory dependent. {{However, the ratio of void statistics in two cosmological models is almost independent of the non-Markov and Markov framework. For halos-in-voids statistics, using the non-Markov framework is complicated and computationally costly. Accordingly, to have an insight into the problem, we compared the two models in the Markovian framework, and we found the specific redshift that we have the maximum difference between the prediction of the two models.}}
We emphasize that this work is a first step in this direction and a proof of concept that halos-in-voids are promising probes for distinguishing different cosmological models. However, comparing the EST results with simulation and observations are much more challenging due to the controversy in literature for finding voids in simulations. Mock catalogs can be used as a first step in studying this concept.
The work can be extended by considering the mass assembly history of halos living in voids, taking into account both mass accretion and mergers.  The effect of the primordial power spectrum's deviation from scale-invariance on void statistics is interesting to examine. To conclude, it is feasible to implement the ideas proposed in this work to distinguish the standard model from alternative cosmological models.\\

\section*{Acknowledgments}
We would like to thank the anonymous referee for his/her insightful comments that elevated the manuscript to a whole new level. Our thanks go out to Mohadese Khoshtinat, Farnik Nikakhtar, and  Arghavan Shafiee for many fruitful discussions.
SB is partially supported by the Abdus Salam International Center of Theoretical Physics (ICTP) under the junior associateship scheme. \\ \\
\section*{Data Availability}
The results are reported based on the data provided by\cite{Wang:2018fng} for the reconstructed Hubble parameter. The data for all plots based on
our theoretical models are available upon request.


\bibliographystyle{mnras}

\begin{twocolumn}
\appendix
\section{Error Propagation Analysis} \label{app:I}
In this section, we study the error propagation of voids and halos-in-voids number density. We calculate and show the result for $\pm 1\sigma$ confidence level in the results and figures. 
We use equation \ref{eq:growth} to calculate growth function $D=D(z)$ with corresponding $\pm 1\sigma$ confidence level (for more discussion, refer to Fig.(1) and Fig.(3) of \cite{Kameli:2020kao}). All other parameters in EST are dependent on growth function. In other words, the cosmological model affects all parameters (critical density, first-crossing, number density) through the growth function. Thus, it is essential to investigate the effect of growth function error propagation on the results. In this section, we intend to show that the effect of growth function error propagation (in our range of interest: radius in voids and halo mass in halos-in-voids) is monotonically increasing.  \\
In the first step, for voids statistics; we study the effect of growth function error propagation on the first-crossing as (equation \ref{eq:ffuvoid}).
\be
{\cal{F}}(S,\delta_v,\delta_c) =\frac{1}{S} \sum_{j=1}^{\infty} \frac{j^2\pi^2{\cal{D}}^2}{\delta_v^2/S}\frac{\sin(j\pi{\cal{D}})}{j\pi}\exp\left(-\frac{j^2\pi^2{\cal{D}}^2}{2{\delta_v}^2/S}\right), \label{eq:app1}
\ee 
which this function depends on growth function through $\delta_{c/v}=\delta_{c0/v0}/D(z)$. Note that ${\cal{D}}\equiv \frac{|\delta_v|}{(\delta_c + |\delta_v| )}={\text{const}}$. Then, we rearrange the equation (\ref{eq:app1}) based on the growth function:
\be
f_{\text{Void}}(D,S) =\sum_{j=1}^{\infty} f_j D^2(z)\exp\left(-g_j(S)D^2(z)\right),
\ee   
where 
\begin{eqnarray}
f_j=\frac{j^2\pi^2{\cal{D}}^2}{\delta_{v0}^2}\frac{\sin(j\pi{\cal{D}})}{j\pi}, \\ \nonumber
g_j(S)=\frac{j^2\pi^2{\cal{D}}^2}{2{\delta_{v0}}^2/S}.
\end{eqnarray}
We show $f_{\text{Void}}$ versus the growth function in Fig. (\ref{fig:app1-ffu-d}) for different void's radii. We show for $R\geq 1.5$ Mpc the $f_{\text{Void}}(D)$ increases monotonically. We use the normalized growth function to unity in the present time, so its value decreases with redshift ($D(z=0.5)\gtrsim 0.8$ and $D(z=1)\gtrsim 0.6$). These results show that in the void's radii range, the void statistics increase monotonically with increasing the growth function. So, larger error propagation in the growth function (introduced by an error in the Hubble parameter) leads to a larger error bar in the results (shaded regions). \\
\begin{figure}
	\includegraphics[width=9cm]{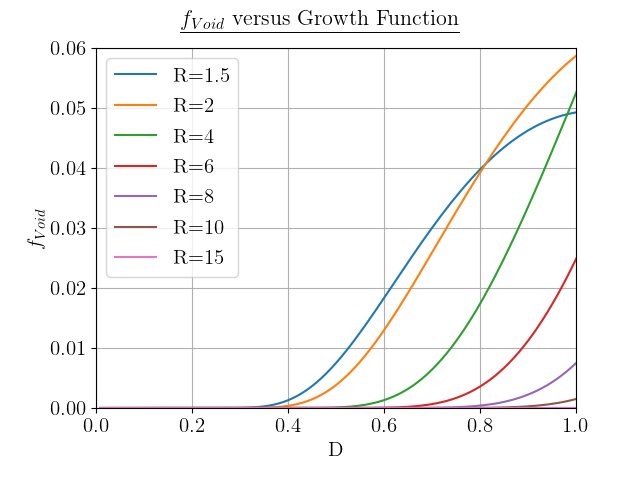}
	\caption{ {{We show the first-crossing function $f_{\text{Void}}$ versus the growth function $D(z)$ in different radii of voids $R >1.5$ .  The $f_{\text{Void}}$ increases monotonically in all ranges of our interest.}}} \label{fig:app1-ffu-d}
\end{figure}
Similarly in the next step, for halos-in-voids statistics, we study the effect of growth function error propagation on conditional first-crossing. (equation \ref{eq:ffuc})
\begin{eqnarray} \label{eq:app2}
f_{\text{FU}}(S_h(M),\delta_c(z)|S_v(R),\delta_v(z))=\frac{1}{\sqrt{2\pi}}.\frac{\delta_c(z) - \delta_v (z)}{(S_h(M)-S_v(R))^{3/2}} \\ \nonumber \times\exp\left({-\frac{(\delta_c(z) - \delta_v(z))^2 }{2(S_h(M)-S_v(R))}}\right),
\end{eqnarray}
Again, we rearrange the conditional crossing equation based on the growth function:
\be
f_{FU}(M|R;D(z))=f(R,M)\frac{1}{D(z)}\exp\left(-g(R,M)\frac{1}{D(z)^2}\right)
\ee
where
\begin{eqnarray}
f(R,M)&=&\frac{1}{\sqrt{2\pi}}.\frac{\delta_{c0} - \delta_{v0}}{(S_h(M)-S_v(R))^{3/2}}, \\  \nonumber
g(R,M)&=&\frac{(\delta_{c0} - \delta_{v0})^2 }{2(S_h(M)-S_v(R))}.
\end{eqnarray}
We calculate ${df_{FU}}/{dD}=0$ to find the conditional first up-crossing maximum as 
\be
D_{max}=\sqrt{2g(R,M)}=\frac{\delta_{c0} - \delta_{v0} }{\sqrt{S_h(M)-S_v(R)}}\simeq \frac{4.4 }{\sqrt{S_h(M)-S_v(R)}}.
\ee 
In the worst scenario for the lowest halo mass ($M=10^{10} M_{\odot}$, with the largest variance $S_h(M)\simeq 11.8$); we calculate the $D_{max}\simeq1.28$ by neglecting the $S_v(R)$. Due to the normalizing growth function, it is obvious that all halo mass range of our interest is monotonically increasing. For larger mass, the maximum of the growth function is larger. For instance for halo mass $M=10^{11} M_{\odot}$ the variance is $S_h(M)\simeq 6.75$ and $D_{max}\simeq 1.69$. Note that $D_{max}=1$ correspond to $M<10^{9} M_{\odot}$ and $D_{max}=0.8 $ correspond to $M<10^{8} M_{\odot}$. Therefore, we conclude that the error propagation for conditional statistics is monotonically increasing for almost the mass range of $M>10^{8} M_{\odot}$.  {{In a future work, one may improve the computational speed of procedure and be able to incorporate the growth function extract from the chains from a cosmological likelihood analysis.}}

\section{Effective Redshift Identification} \label{app:II}
We show the void number density and the number density ratio of the two cosmological models in two arbitrary void radii in Fig. (\ref{fig:nVoid-z}) with their $\pm 1 \sigma$ error propagation. We found the effective redshift is  $z\sim 1.7$. This redshift corresponds to the maximum deviation for the voids number density of the two cosmological models. For a more comprehensive study, we show the same result in Fig. (\ref{fig:app2-void}) for more various voids. As shown in this figure, the effective redshift remains the same for all different radii, and the ratio magnitude increases by increasing the void's radius. \\
\begin{figure}
	\includegraphics[width=9 cm]{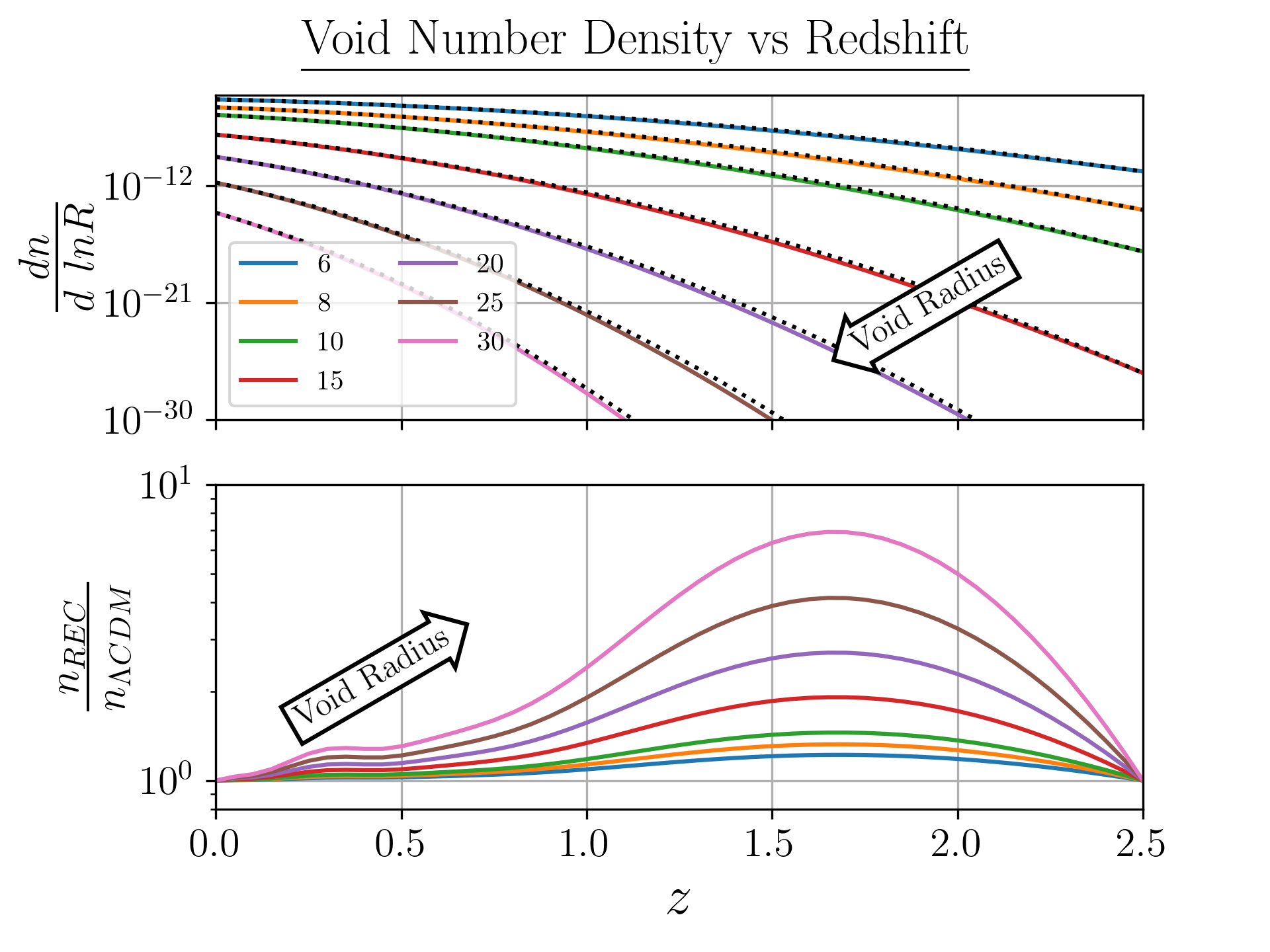}
	\caption{The number density of voids are plotted versus redshift for various radii $R=6,8,10,15,20,25,30$ Mpc. The upper panel shows the prediction of the two cosmological models ($\Lambda$CDM (solid lines) and Rec (dotted lines)), and the bottom panel shows the ratio of the two models. This figure is complimentary to Fig. (\ref{fig:nVoid-z}) in the main text.} \label{fig:app2-void}
\end{figure}
We show the halos-in-voids number density and the cosmological model ratio for two arbitrary halo masses for two void's radii in Fig (\ref{fig7}). The effective redshift for the conditional statistics is $z\sim 1.9$. For a more detailed study, we show the result for different halo masses in Fig. (\ref{fig:app2-haloinvoid}). As depicted in this figure, the effective redshift remains the same. Furthermore, the ratio of number density increases for larger halo masses. \\    
\begin{figure}
	\includegraphics[width=9cm]{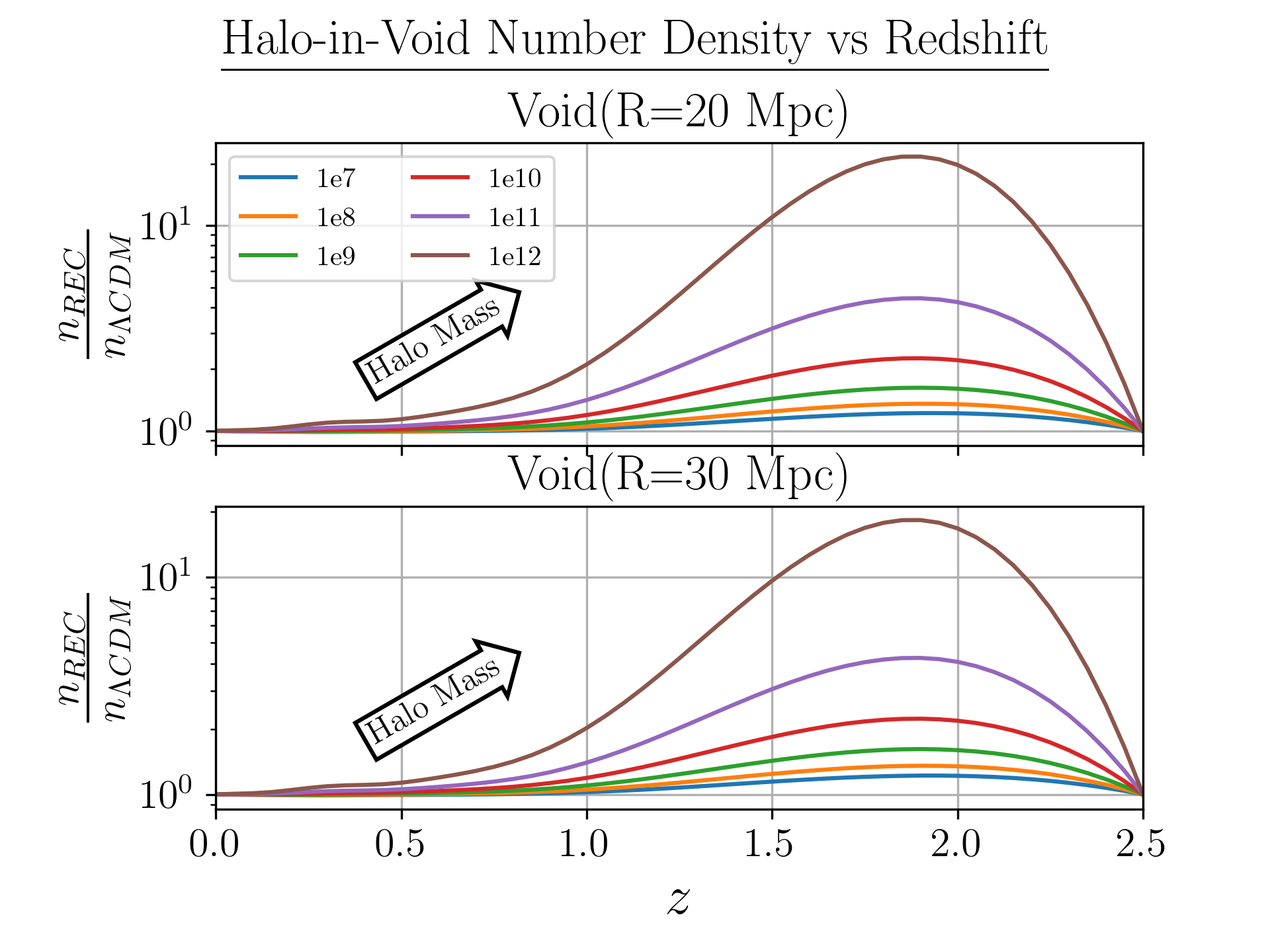}
	\caption{The number density of halos-in-voids are plotted versus redshift for  DM halo masses $M=10^{7},10^{8},10^{9},10^{10},10^{11}, 10^{12} M_{\odot}$. The prediction of the two cosmological models ($\Lambda$CDM and Reconstructed plotted for host void with $R=20$Mpc radius in the upper panel and for $R=30$ Mpc in the bottom panel. This figure is complimentary to Fig. (\ref{fig7}) in the main text.} \label{fig:app2-haloinvoid}
\end{figure}


\bsp	
\label{lastpage}
\end{twocolumn}
\end{document}